# Theory and Experiments of Pressure-Tunable Broadband Light Emission from Self-Trapped Excitons in Metal Halide Crystals

Shenyu Dai[1,2,†] , Xinxin Xing[2,3,†], Viktor G. Hadjiev[4,5,*], Zhaojun Qin[2,3], Tian Tong[2], Guang Yang[6], Chong Wang[2,7], Lijuan Hou[2], Liangzi Deng[4], Zhiming Wang[3], Guoying Feng[1,*] and Jiming Bao[2,4,6,*]

[1]College of Electronics & Information Engineering, Sichuan University, Chengdu, Sichuan 610064, China

[2]Department of Electrical and Computer Engineering and Texas Center for Superconductivity (TcSUH), University of Houston, Houston, Texas 77204, USA

[3]Institute of Fundamental and Frontier Sciences, University of Electronic Science and Technology of China, Chengdu, Sichuan 610054, China

[4]Department of Physics and Texas Center for Superconductivity (TcSUH), University of Houston, Houston, TX 77204, USA

[5]Department of Mechanical Engineering, University of Houston, Houston, TX 77204, USA

[6]Materials Science & Engineering, University of Houston, Houston, Texas 77204, USA

[7]International Joint Research Center for Optoelectronic and Energy Materials, School of Material Science and Engineering, Yunnan University, Kunming, Yunnan 650091, China

[†]These authors contributed to the work equally

*To whom correspondence should be addressed: Jiming Bao (jbao@uh.edu ), Guoying Feng (guoing_feng@scu.edu.cn), Viktor Hadjiev (vhadjiev@uh.edu )

# Abstract

Hydrostatic pressure has been commonly applied to tune broadband light emissions from self-trapped excitons (STE) in perovskites for producing white light and study of basic electron-phonon interactions. However, a general theory is still lacking to understand pressure-driven evolution of STE emissions. In this work we first identify a theoretical model that predicts the effect of hydrostatic pressure on STE emission spectrum, we then report the observation of extremely broadband photoluminescence emission and its wide pressure spectral tuning in 2D indirect bandgap $CsPb_2Br_5$ crystals. An excellent agreement is found between the theory and experiment on the peculiar experimental observation of STE emission with a nearly constant spectral bandwidth but linearly increasing energy with pressure below 2 GPa. Further analysis by the theory and experiment under higher pressure reveals that two types of STE are involved and respond differently to external pressure. We subsequently survey published STE emissions and discovered that most of them show a spectral blue-shift under pressure, as predicted by the theory. The identification of an appropriate theoretical model and its application to STE emission through the coordinate configuration diagram paves the way for engineering the STE emission and basic understanding of electron-phonon interaction.

## Introduction

Broadband light emission from self-trapped exciton (STE) in metal halide perovskites has attracted enormous attention because it can not only be used for applications such as single-component white light sources, but also provides a unique venue for studying strong electron-phonon interactions(*1-5*). Several reports demonstrate significant spectral tuning and intensity changes of STE emission under pressure(*6-11*). However, a critical and general understanding of the effect of hydrostatic pressure on STE emission is missing. Some simply attributed it to pressure-induced effect without providing concrete understanding, while others invoked coordination configuration diagrams, but used diagrams arbitrary for cartoon illustration of their observations. There are a few first-principles calculations (FPC) of STEs in metal halide perovskites(*12-14*). These calculations, however, are difficult to be extended to STE under pressure in part because FPC usually fail to reproduce a non-monotonous bond length change with pressure as that observed in $CsPb_2Br_5$(*15*). Although providing insightful results, the FPC lacks the functionality of some of the parametric models that convey more transparently the pressure action on STEs. So far, there is no theoretical model that has been used to predict whether the STE spectrum in perovskites will shift to higher or lower energy under pressure. The lack of functional models prevents us from gaining insights into electron-phonon interactions and hinder potential applications of STE emission.

In this work, we first identify a theoretical model based on the standard coordinate configuration diagram that correctly describes the effect of pressure on the STE absorption line. We then present our original observations of extremely broad STE light emission in the indirect wide-bandgap $CsPb_2Br_5$ crystals and the ultra-wide spectral tuning of this emission under moderate pressure. After identifying the free exciton (FE) emission of $CsPb_2Br_5$ in the low temperature photoluminescence (PL) spectra and having confirmed it by first-principles calculations, we establish a proper configuration coordinate diagram that provides a notable agreement between the theory and our experiment: a linear increase in STE energy with pressure and a nearly pressure-independent STE emission linewidth. Based on the model and experimental results at higher pressure, we show that at least two groups of STEs coupled to multiple phonons are involved in the STE PL emission. At the end, we review recent publications on pressure evolution of STE emission and find that nearly all of them exhibit a spectral blue-shift, in agreement with the model presented in this paper.

## Modeling of STE under pressure

The theory of STE has been well elaborated (*16, 17*), but the effect of pressure on STE emission has remained unsettled. Fig. 1a shows the standard configurational coordinate diagram that plots the adiabatic potential energy surfaces (APES) of free exciton (FE) and STE along the configurational coordinate $Q$ of atomic displacements. We emphasize the importance of approximations on which an APES model is build – something that is largely missing in recent perovskite literature. Here based on the adiabatic Bohr-Oppenheim (BO) approximation and the linear electron-phonon coupling Frank-Condon (FC) approximation, the lattice relaxation energy is related to the phonon creating the localization force field in the ground $g$ and excited states $e$ as $E_{RL}^{g(e)} = S_{g(e)}\hbar\omega_{g(e)}$. In this expression, $S_{g(e)}$ are the Huang-Rhys (HR) factors that represent the average number of phonons created during a vertical absorption $S_e$ and emission $S_g$ transitions. HR factors are a measure for the strength of electron-phonon coupling with phonons $\omega_g$ and $\omega_e$ in the ground and excited states. Generally, the phonon mode frequencies in the ground and excited states are not equal but in some cases they can be very close. The HR factor in the ground state also determines the PL linewidth of STE through $\Gamma_{STE} = 2.36\sqrt{S_g}\hbar\omega_g$ at low temperatures(*18*). The energy of STE PL band center in Fig. 1(a), $E_{STE}$, is given by $E_{STE} = E_{FE} - E_{RL}^{g} - E_{ST}$. The energy barrier $E_{bar}^{FE}$ separates the FE from STE states, and $E_{bar}^{NR}$ is the energy needed for a transition from radiative to multi-phonon non-radiative recombination of STE.

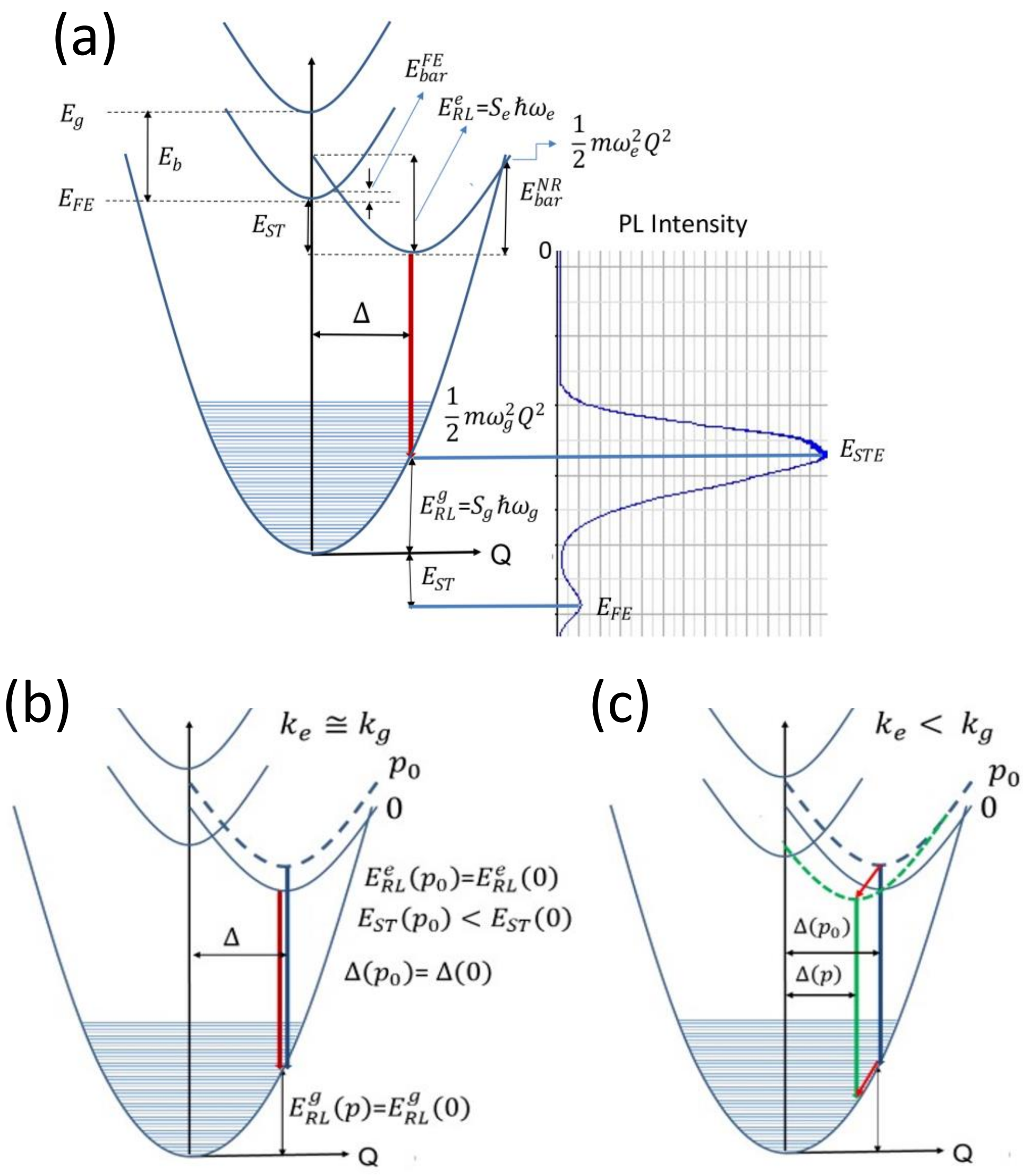

Figure 1. (a) Configuration coordinate diagram (exaggerated drawing) of one-phonon, single self-trapped exciton (STE). $E_g$ is the exciton ionization (electronic bandgap) energy, $E_b$ – the FE binding energy, $E_{FE}$ is the free exciton energy, $E_{bar}^{FE}$ is the energy barriers between STE and FE states. $E_{bar}^{NR}$ is the energy needed for a STE to relax through multi-phonon non-radiative recombination. Evolution of the configuration diagram with pressure: (b) $k_e \cong k_g$ and (c) $k_e < k_g$.

The evolution of a configurational diagram, as that in Fig. 1a, under hydrostatic pressure and the corresponding optical absorption were studied in detail in Refs.(*19-21*) under simplified assumptions of no change of vibrational frequencies with pressure. Here, we extend these studies to emission processes and accounting for the pressure dependence of vibrational frequencies. For details see the Supplemental Information (SI). Since the potential vibrational energy in the ground and excited states is that of harmonic oscillators, the problem of configurational diagram modification under hydrostatic pressure is essentially that of the forced harmonic oscillators. It is well known that applying a constant force to a harmonic oscillator leads only to displacement of the equilibrium position (minimum) of its potential energy well, but for a system of oscillators their relative potential energies change as well. Under the approximation in which $Q$ displaces linearly with pressure $p$, the change in the STE PL band position $E_{STE}(p)$ at pressure $p$ with respect to ambient, $p = 0$, $E_{STE}(0)$ is given by (*20*):

$$E_{STE}(p) = E_{STE}(0) + \left(1 - \frac{\omega_g^2(p)}{\omega_g^2(0)}\right) E_{RL}^{g}(0) + \Delta(0)\frac{k_g}{k_e}p + \frac{1}{2}\frac{(k_e - k_g)}{k_e^2}p^2, \qquad (1)$$

where $k_{g(e)} = m\omega_{g(e)}^2(p)$ are the force constants of participating phonons in the ground $g$ and excited $e$ states at pressure $p$, $m -$ masses of vibrating atoms, and $\Delta(0)$ is the $Q$-distance between the minima of $g$ and $e$ potential wells at $p = 0$, denoted as Δ in Fig. 1(a). $E_{RL}^{g}(0)$ is the relaxation energy at zero external pressure. Under pressure $p$, Δ changes as

$$\Delta(p) = \Delta(0) + \frac{(k_e - k_g)}{k_g k_e}p, \qquad (2)$$

and the STE PL linewidth also depends on $p$:

$$\frac{\Gamma_{STE}(p)}{\Gamma_{STE}(0)} = \left(\frac{\omega_g(p)}{\omega_g(0)}\right)^{3/2} \frac{\Delta(p)}{\Delta(0)} \ . \qquad (3)$$

The second term in Eq. 1 represents one of the effects of increasing $\omega_g(p)$ with pressure, namely the curvature of the ground state potential energy well versus $Q$ increases and this alone results in decreasing $E_{STE}(p)$. We especially note this effect because it is easily predictable but can produce only red shifted $E_{STE}(p)$ with increasing pressure. The blue-shifted STE PL comes from the last two terms in Eq. 1 when they prevail over the second one.

In the simplest case, $k_e \cong k_g$ ($\omega_g \cong \omega_e$) and a weak dependence of the phonon frequency on pressure $\omega_g(p) \cong \omega_g(0)$, shown in Fig. 1b, the STE PL band energy increases linearly with pressure due to decreasing of the self-trapped energy $E_{ST}$, while both $\Delta$ and the PL linewidth remain constant as follows from Eqs. (1) – (3). Although these results are somewhat surprising, they comply with the so-called rigid shift theorem stating that an optical absorption line under pressure shifts in energy without change in shape if $\omega_g = \omega_e$(*22*). Fig. 1c presents the case, in which at pressure $p > p_0$, the trapping phonon frequencies in the ground and excited states start to deviate and $k_e < k_g$.

## Experimental Results

Large size $CsPb_2Br_5$ microplates were synthesized simply by dropping $CsPbBr_3$ micro-powder in large quantities of deionized water in a flask (*23, 24*) (Fig. 2a-b). Due to the indirect bandgap of 3.35 eV, these crystals are transparent and produce no light emission under the sub-bandgap 473 nm (2.62 eV) laser excitation (Fig. 2c). They appear, however, very differently under UV illumination. Figure 2b shows that $CsPb_2Br_5$ microplates emit red light under a 364 nm (3.41 eV) UV LED. Fig. 2c shows that the light emission is very broad with a full width at half maximum (FWHM) of 675 meV. Note that the seemingly edge emission in Fig. 2b is due to wave guiding of emission inside thick slabs of transparent microplates. Fig. 2d shows that the emission intensity is proportional to the incident laser power. Based on the above observed dependence of PL on the excitation wavelength and power, we can exclude defect states as the origin of broadband light emission(*15, 23*) and explore it as emission from STE. Fig. 2e shows PL at low temperatures, a peak around 423 nm emerges at ~130 K and becomes stronger in intensity with lowering temperature. Fig. 2f plots the intensity of 423 nm PL line as a function of temperature. The temperature dependence of 423 nm PL intensity is characterized with 10 meV activation energy obtained from the fit shown in Fig. 2f. We assign the 423 nm PL peak to free exciton (FE) emission and the broadband to STE emission. Given that FE PL emission in $CsPb_2Br_5$ is observed only below 130K, we conclude that the corresponding 10 meV activation energy is $E_{bar}^{FE}$ in Fig. 1a - the barrier energy for thermally excited transitions from free to self-trapped excitons.

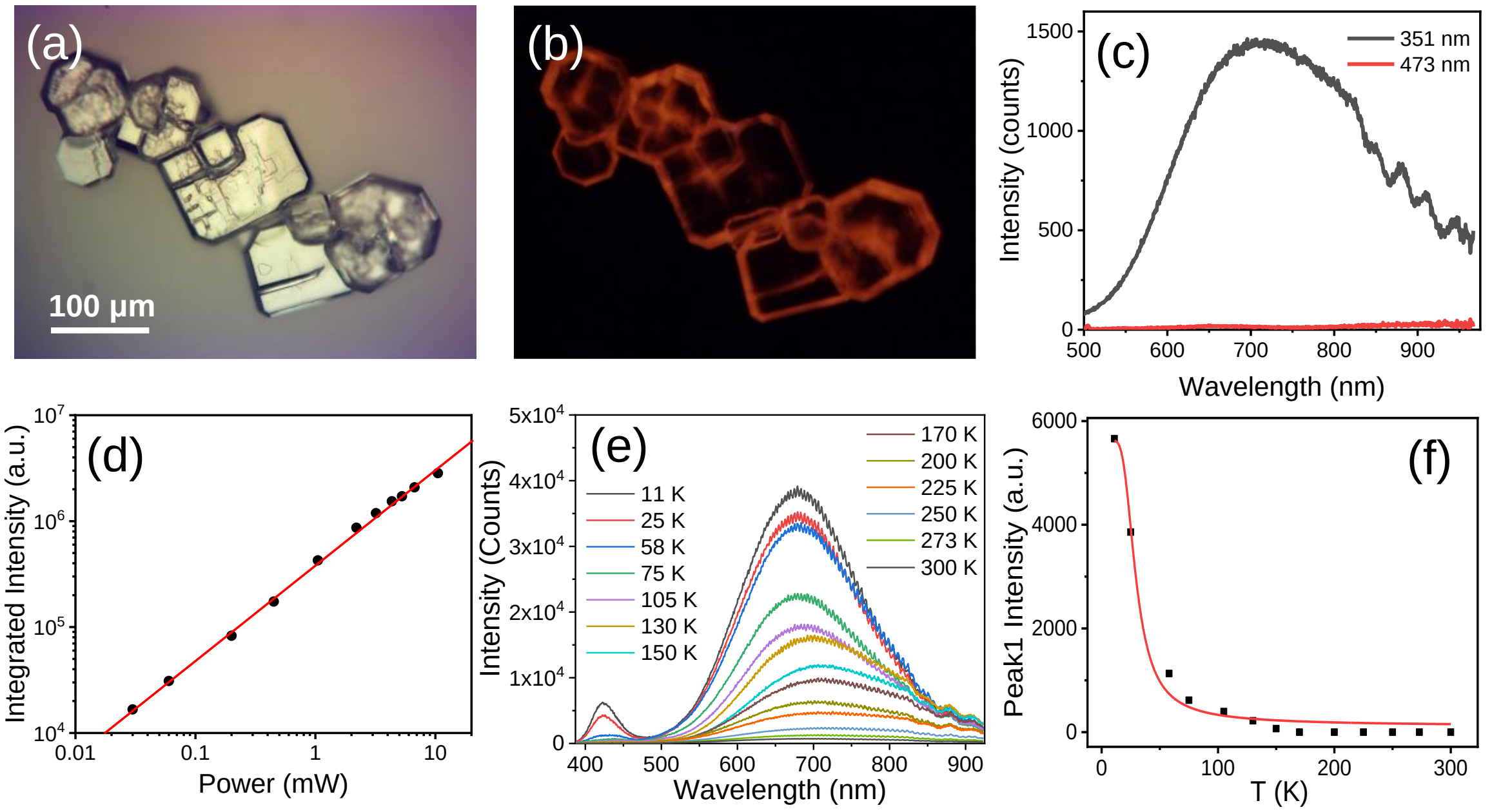

Fig. 2. $CsPb_2Br_5$ crystals and their broadband photoluminescence spectra. (a-b) Optical and photoluminescence image of large size microplates under (a) white light and (b) 364 nm (3.41 eV) UV light. (c) Photoluminescence of a microplate excited by 351 nm (3.53 eV) and 473-nm lasers. (d) Photoluminescence intensity of the same $CsPb_2Br_5$ microplate as a function of 351-nm incident laser power. (e) Temperature-dependent photoluminescence spectra at ambient pressure. (f) Intensity of free exciton emission as a function of temperature from (e).

Our previous study shows that hydrostatic pressure is very effective to distinguish STE emission from defect emission and tune the emission peak and intensity(*23, 25-27*), thus we applied the same diamond anvil cell (DAC) to $CsPb_2Br_5$. Fig. 3a shows a picture of a $CsPb_2Br_5$ microplate loaded in the cell. Fig. 3b shows that from 0 to 1.63 GPa, the PL intensity increases significantly. At the same time, the PL peak experiences a large blueshift. After that, the intensity starts to decrease, while the broad light emission band continues to shift to shorter wavelength, and eventually disappears at above 5 GPa. The large PL bandwidth accompanied with the blue shifted band center makes it possible for the broad light emission to cover a wide spectral range from near-infrared to blue. This also can be seen in the direct optical images in Fig. 3c. In particular, a warm bright white light is produced near 1.63 GPa, which is surprising and significant for an indirect bandgap crystal.

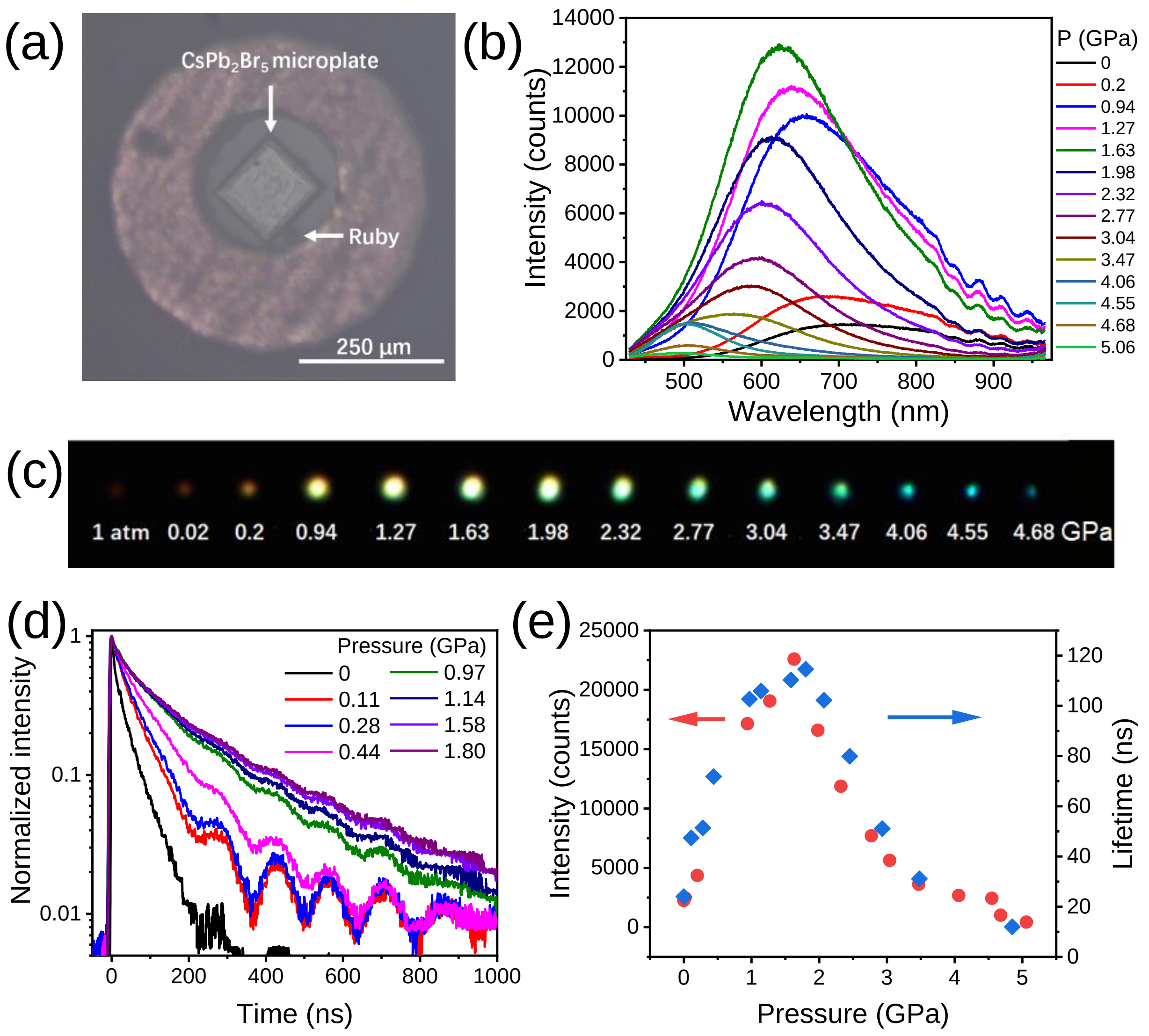

Figure 3. Effect of pressure on STE light emission. (a) Optical micrograph of a $CsPb_2Br_5$ microplate loaded in the DAC. (b) Evolution of PL spectrum of the $CsPb_2Br_5$ microplate under hydrostatic pressure. (c) Optical images of broadband light emission under increasing pressure. (d) Time-resolved PL at ambient temperature. (e) PL lifetime and intensity as a function of pressure.

To have a better understanding of the pressure dependent PL, we also performed time-resolved photoluminescence, Raman and UV-vis spectroscopies under various pressures. Fig. 3d shows a representative PL decay curve of the $CsPb_2Br_5$ microplate at ambient temperature and Fig. 3e summarizes the change of decay time with pressure. In the same figure we compare the pressure-dependent decay time with the corresponding PL intensity variation and demonstrate they are correlated. The Raman results agree with our previously reported observations (*23, 24*) and are included in the Supplementary Online Materials. Due to the lattice compression, the Raman spectrum experiences an overall blueshift, and notably the $A_{1g}$ phonon at 81 cm$^{-1}$ begins to separate from the nearly overlapping $B_{1g}$ phonon for pressures above 2 GPa (*23*). There is no change in the optical absorption edge under pressure below 2 GPa, and only a small redshift of ~20 meV from 2 GPa to 5 GPa is observed (Fig. S4 Supplementary Online Materials). These findings indicate that the applied pressure induces only elastic deformations of the lattice and no phase transitions and significant changes in the electronic and phonon properties are observed.

## Discussions

Before comparing the theory with our experiments, we want to further prove that the 423 nm low temperature emission is due to FE radiative recombination as this is an essential parameter in the STE theory in Fig. 1. We performed the density functional theory (DFT) calculations with Heyd-Scuseria-Ernzerhof (HSE) functional and allowing for the spin-orbit coupling (SOC) (*28*). Fig. 4a shows the electronic band structure of $CsPb_2Br_5$. The indirect bandgap with energy 3.45 eV is between N and Γ points in the Brillouin zone (BZ) of the primitive cell depicted in Fig. 4c. Fig. 4b (blue curve) shows the optical spectra of $CsPb_2Br_5$ calculated within the approximation of independent particles (IP). The spectrum corresponds to the optical bandgap edge absorption without the excitonic effects and transitions across the indirect bandgap. However, the calculations at the many-body perturbation level of theory by solving Bethe-Salpeter equation (BSE) (*29*) produce two exciton lines at 2.96 eV and 3.11 eV for the Γ-Γ optical transitions (the lowest energy direct bandgap) that are close in energy to the experimental FE PL band. Fig. 4d plots the exciton spatial distribution as the probability of finding the excited electron around a fixed hole. It is clearly seen in the upper panel of Fig. 4d that the exciton is confined within a single Pb-Br layer, that is, it is essentially a 2D exciton with 10 Å in-plane radius. This result is also reproduced by a 2D Wannier-Mott model as shown in the Supplementary Online Materials. The analysis of DFT and BSE calculations reveals that the holes in the direct bandgap excitons reside on Br atoms (4*p* states) and the excited electrons are around Pb atoms (6*p* states).

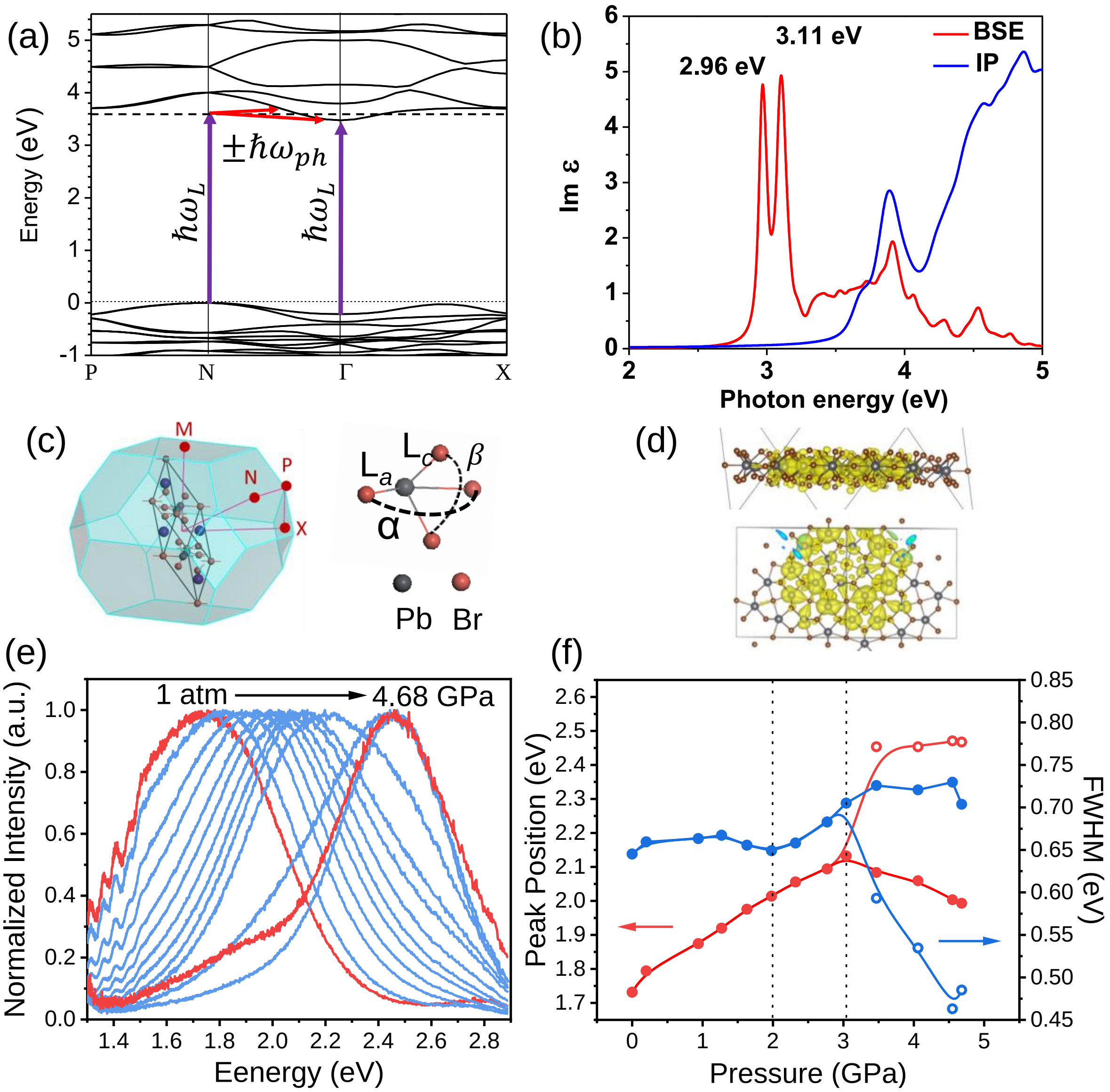

Figure 4. DFT simulation of free exciton emission and effect of pressure on spectrally resolved STE light emission. (a) Band structure of $CsPb_2Br_5$ with the calculated indirect N-Γ and lowest direct bandgap Γ-Γ transitions at 3.45 eV (exp. 3.35 eV) and 3.64 eV, respectively. The laser excitation 3.53 eV is scaled-up to 3.63 eV to compensate the overestimated calculated energies. (b) Calculated imaginary part of macroscopic dielectric function (optical absorption) at the BSE level of theory (red curve) and at independent particle (IP) approximation (blue curve). (c) Brillouin zone (BZ) and primitive cell of $CsPb_2Br_5$ crystal. (d) Two projections of spatial distribution of the exciton wavefunction showing 2D exciton with radius ~10 Å. The upper panel is in the plane of c-axis, the lower one – normal to the c-axis (e) Normalized STE PL lineshape evolution with pressure. (f) PL band

positions and FWHM of $CsPb_2Br_5$ versus pressure.

To obtain a better view of the PL line shape evolution with pressure, we present in Fig. 4e the normalized experimental spectra. We note that the PL band cannot be fitted well enough with a single Gaussian lineshape as predicted by one-phonon, single-STE models (*30*). A splitting of the PL band into two sub-bands, however, becomes notable for pressures higher than 3 GPa and the corresponding PL band fitting results are presented in Fig. 4f. Given the splitting of the PL at pressure $>$ 3 GPa, we believe that at lower pressure the two PL bands nearly overlap and stem from at least two different STEs coupled to phonons with similar pressure dependence of their frequencies. Apparently, the pressure dependence of STE PL is very complex and at this stage of study it would be very instructive if we can have a qualitative understanding of the experimental observation.

The pressure range of 0-2 GPa is particularly interesting because the energy bandgap changes are negligibly small and the same can be expected for the free exciton energy $E_{FE}$ (*23*). As seen in Fig. 4e, in this pressure range the PL bandwidth remains almost unchanged, whereas PL peak position shifts linearly to higher energies with pressure. These results are in surprising accordance with the prediction of Eqs. (1) – (3) and Fig. 1(b) for $k_e \cong k_g$ of a phonon field with slight to no change of phonon frequency $\omega_g(p) \cong \omega_g(0)$ with pressure. The pressure effect comes from the increasing of energy difference between the minima (equilibrium positions) of the excited and ground potential wells with pressure while preserving $\Delta$ and the corresponding HR factors as follows from the expression $E_{RL}^{g} = S_g \hbar\omega_g = \frac{1}{2} m\omega_g^2 \Delta^2$. Figs. S2-3 and our previous studies show that two of the Raman phonons change very little their frequency with pressure in the range 0-2 GPa (*23, 24*). These are the $A_{1g}$ phonon at 133 $cm^{-1}$ involving Br - Pb stretching of $L_c$ bond length in Fig. 4c and the $B_{2g}$ symmetry Pb vibrations at 69 $cm^{-1}$ that corresponds to out-of-phase Br-Pb-Br bending. We believe that these or/and other non-Raman active phonons that have similar pressure dependence are creating the phonon self-trapping field of the two STEs.

In the transitional pressure range 2 GPa to 3 GPa, the PL bandwidth starts to increase while the PL band is still blue shifted. We relate this behavior to the inhomogeneous broadening as the PL energies of the two STEs start to change at different pace with pressure. Notably, a collapse of the c-axis of $CsPb_2Br_5$ crystals under pressure was observed around 2 GPa (*15*). This reversible isostructural phase transition is expected to have greater impact on the $A_{1g}$ mode at 133 $cm^{-1}$ than that of the Pb $B_{2g}$ mode at 69 $cm^{-1}$. Indeed, with further increasing the pressure, the $A_{1g}$ mode frequency begins to increase rapidly, whereas the pace of frequency change of $B_{2g}$ mode remains low.

Above the apparent split at ~3 GPa, the energy of one of the PL bands, empty circles in Fig. 4f, remain nearly constant and the other (solid circles) start to decrease slightly. We believe that in this pressure range the phonon frequencies in the excited state begin to deviate from those in the ground state. As a result, the term proportional to $p^2$ in Eq. 1 begins to compensate the changes due to the linear in pressure term, provided $k_e < k_g$, as demonstrated in Fig. 1c. Under same conditions, according to Eqs. (2) and (3) the PL bandwidth should decrease, which is also seen in the experimental results (empty circles) in Fig. 4f. We suggest that the empty circle branch corresponds to the energy and linewidth of the STE involving the slow frequency changing with pressure $B_{2g}$ mode. The full circle graphs are consistent with the $A_{1g}$ mode coupled STE. The increase of $A_{1g}$ frequency with pressure works towards decreasing the STE energy due to the second term in Eq. 1 and the difference between $k_e \neq k_g$ brings additional subtleness of the STE behavior. While the model behind Eqs. (1)-(3) explains well the PL pressure dependence, it is not suitable enough to deliver quantitative results. For instance, the quantity $p$ called pressure in the model has dimensionality of a force. It is related to the external pressure $P$ as $p = P.A$, where $A$ is the coupling constant of $P$ to the STE (*19, 21*). The quantity $A$ does not depend on pressure, has dimensionality of area, but cannot be determined independently.

The diagram in Fig. 1b and Eqs. (1)-(3) can also explain enhanced PL intensity and lifetime under initial mild pressure. As discussed above, when there is no noticeable change to both electronic and lattice structures under such a mild pressure, both the radiative and conventional non-radiative recombination rate should stay the same. The nature of STE has provided extra channels to control the dynamics and emission of excitons. The population and recombination of STE depend on two potential barriers: $E_{bar}^{NR}$, which will prevent non-radiative recombination through the crossing between STE and the ground state; $E_{bar}^{FE}$, which will prevent backward transition from STE to free exciton. Initial pressure will move STE upwards and result in increased PL energy. It will also make $E_{bar}^{NR}$ higher and thus reduce non-radiative recombination, leading to increased PL lifetime and enhanced PL intensity. However, more upward displacement of STE with increasing pressure will make it easier for STE to return back to free exciton state, leading to reduced PL lifetime and intensity.

Since the model presented in this paper has explained most of the experimental results, we use it to assess the pressure effects on STE in other perovskites. We want to emphasize that the model can reproduce any direction of the spectral shift of STE PL depending on the combination of the pressure-dependent participating phonon frequencies and their difference in the ground and excited states. We have found that the significant spectral blue-shift of STE emission under pressure is the most common phenomenon (*6-11*), and it has been observed in 0D hybrid halide $(C_5H_7N_2)_2ZnBr_4$ (*8*), 0D lead-free perovskite $Rb_7Sb_3Cl_{16}$(*11*), 1D metal halide $C_4N_2H_{14}PbBr_4$(*7*), double perovskite such as $Cs_2AgBiCl_6$, Bi-doped $Cs_2AgInCl_6$, and $Cs_2AgInCl_6$ (*6, 9, 10*). Two reports show very little or no obvious blue-shift (*31, 32*). For example, there is almost no peak shift in corrugated 1D hybrid metal halide $C_5N_2H_{16}Pb_2Br_6$ from 0-3 GPa, while at the same time, the bandgap is reduced by 160 meV (*31*). A possible reason for that is the self-compensating effect of a pressure increase of phonon frequencies resulting in STE energy decrease that works against the linear in pressure term in Eq. 1. Surprisingly, we have found no report on the opposite effect, i.e., spectral red shift of STE with pressure although the bandgap of most metal halides decrease under pressure. We relate this observation to the dominance of always active and increasing with pressure term $\Delta(0)\frac{k_g}{k_e}p$, while the other pressure dependent terms in Eq. 1 can be zeroed.

## Conclusion

In summary, we have identified a STE theoretical model and adopted it to understand evolution of STE emission in 2D $CsPb_2Br_5$ crystals under hydrostatic pressure. An excellent agreement between the model and experiments is obtained regarding the effect of pressure on STE PL peak position and peak linewidth. A careful analysis of the experimental results by using the model also reveals that two STEs are involved in the ultra-wide STE emission of $CsPb_2Br_5$. The model prediction of blue-shifted with pressure STE PL peak is also confirmed by many recently reported experiments in a variety of perovskites. This is the first model based on a general theory that help us to understand the effect of pressure and correctly predicted a major spectral change. More pressure-induced changes can be understood if more information such as specific phonons and their excited state phonon frequency is obtained. Hydrostatic pressure has been widely used to tune the STE emission, a general model will certainly help us to better design and engineer the STE for high efficiency broadband emission.

## Supplementary Online Materials

Theoretical modelling, DFT, Calculation of free exciton radius, Pressure evolution of Raman and absorption spectra.

## ACKNOWLEDGMENTS

J.M.B. acknowledges support from the Welch Foundation (E-1728), and a UH Small Equipment Grant (000182016).

## AUTHOR DECLARATIONS

Conflict of Interest

The authors have no conflicts to disclose.